\documentclass[%
,secnumarabic%
,amssymb, amsmath,nobibnotes, aps,nofootinbib,showpacs,]{revtex4}
\usepackage{epsfig}%
\usepackage{graphicx}%
\expandafter\ifx\csname package@font\endcsname\relax\else
 \expandafter\expandafter
 \expandafter\usepackage
 \expandafter\expandafter
 \expandafter{\csname package@font\endcsname}%
\fi

\begin{document}

\title{The Hawking-Unruh phenomenon on graphene}
\author{\it Alfredo Iorio$^1$\footnote{Corresponding author. Tel: +420 22191 2447; Fax: +420 22191 2434; E-mail: iorio@ipnp.troja.mff.cuni.cz}\& Gaetano Lambiase$^{2}$}
\affiliation{$^1$Faculty of Mathematics and Physics, Charles University \\
V Hole\v{s}ovi\v{c}k\'{a}ch 2, 180 00 - Prague 8, Czech Republic.}
\affiliation{$^2$Department of Physics ``Caianiello'', University of Salerno \\
Via Ponte Don Melillo, 84084 - Fisciano (SA), Italy \\
INFN, Sezione di Napoli, Italy.}

\date{\today}

\begin{abstract}
We find that, for a very specific shape of a monolayer graphene sample, a general relativistic-like description of a back-ground spacetime for graphene's conductivity electrons is very natural. The corresponding electronic local density of states is of finite temperature. This is a Hawking-Unruh effect that we propose to detect through an experiment with a Scanning Tunneling Microscope.
\end{abstract}


\maketitle

\bigskip

Graphene is an allotrope of carbon that was first theoretically posited\cite{wallace} and then found to have an abundance of ``unorthodox'' properties\cite{geimnovoselovFIRST}, the understanding of which is appealing to condensed matter as well as high energy theorists, see, e.g., \cite{geim}. In this letter we show that graphene can serve as a realization of the Hawking-Unruh effect\cite{hawking,unruh}, namely of the most crucial prediction of quantum field theory (QFT) in curved spacetimes\cite{birrell,takagi}. This effect (that is eluding direct observations since its proposal, nearly forty years ago), predicts that quantum fields in a spacetime with an horizon exhibit a thermal character due to the nature of the quantum vacuum and to the relativistic process of measurement. This is the first step towards a quantum theory of gravity and, as such, drives a huge amount of constantly ongoing research. The results of this letter are also timely for the current efforts of the condensed matter theory community, as a central issue in the ongoing studies of graphene is how the curvature of the sample modifies its electronic properties\cite{conference2011}.

As is by now well known, the special topology of graphene's Honeycomb lattice (two interpenetrating triangular lattices) is the reason of the effectiveness of the description of its electronic properties in terms of massless, neutral, (2+1)-dimensional, Dirac pseudoparticles \cite{pacoreview2009}. Linearizing around the two inequivalent Fermi points (Dirac points), $\vec{k}^D_\pm$, $\vec{k}_\pm \simeq \vec{k}^D_\pm + \vec{p}$, the tight-binding Hamiltonian can be written as ($\hbar = 1$) \cite{pacoreview2009} $H = v_F \sum_{\vec{p}} \left(\psi_+^\dagger \vec{\sigma} \cdot \vec{p} \; \psi_+  + \psi_-^\dagger \vec{\sigma}^* \cdot \vec{p} \; \psi_- \right)$, or, in configuration space and in the continuum approximation
\[
H = - i v_F \int d^2 x \left( \psi_+^\dagger \vec{\sigma} \cdot \vec{\partial} \; \psi_+
    + \psi_-^\dagger \vec{\sigma}^* \cdot \vec{\partial} \; \psi_- \right)\;,
\]
where $\vec{\sigma} \equiv (\sigma_1, \sigma_2)$, $\vec{\sigma}^* \equiv (-\sigma_1, \sigma_2)$, $\sigma_i$ are the Pauli matrices, $v_F \equiv 3 \eta \ell / 2$ is the Fermi velocity (that will be set to 1) with $ \eta \simeq 2.7$~eV  the hopping parameter and $\ell \simeq 2.5${\AA} the lattice spacing, and $\psi_+^T \equiv (a_+\,\, b_+)$, $\psi_-^T \equiv (a_- \,\, b_-)$ are two-component Dirac spinors, as appropriate for this 2+1-dimensional system ($a$ and $b$ are anti-commuting annihilation operators for an electron in the two sub-lattices). We do not consider short range scattering centres or any other effect mixing the two Fermi points, thus we discuss the physics around a {\it single} Fermi point, e.g. $\psi \equiv \psi_+$. The corresponding action is $A =   i \int d^3 x  \bar{\psi} \gamma^a \partial_a \; \psi$, where $\gamma^0 = \sigma_3$, $\gamma^1 = i \sigma_2$,
$\gamma^2 = - i \sigma_1$ which obey $[\gamma^a , \gamma^b ]_+ = 2 \eta^{a b}$, with $a,b = 0,1,2$ the Lorentz/flat indices (on indices and other geometric conventions see \cite{note1}).

Following the spirit of the condensed matter analogues of gravitational effects\cite{volovik}, and paying due attention to the 2+1 dimensions\cite{jackiw} and to the Weyl symmetry of the massless Dirac field description\cite{iorio}, in this work we shall use graphene as a physical realization of QFT in curved spacetimes. We shall identify a specific shape, the Beltrami pseudosphere, for which it is easier to probe whether graphene quasi-particles experience a general relativistic-like spacetime, hence, as a result of QFT in curved spacetimes, give rise to a thermal spectrum in the form of a finite temperature electronic local density of states (LDOS). The temperature is of the Hawking-Unruh type \cite{hawking,unruh}, depends upon the curvature and the meridian coordinate of the surface, and we show here how to measure it in a dedicated experiment with a Scanning Tunneling Microscope (STM). Noticeably, due to the odd dimensions, the formula contains a Bose-Einstein spectrum for this Dirac system\cite{takagi,hyun1}.

For graphene, to include time in a relativistic fashion we need to take two crucial steps\cite{iorio}. First, we have to assume that for a curved graphene sample the conductivity electrons experience a spacetime metric, the choice dictated by everyday practice being
\begin{equation}\label{mainmetric}
g^{\rm graphene}_{\mu \nu}  (q) = \left(\begin{array}{cc} 1 & 0  \quad 0 \\ \begin{array}{c} 0 \\ 0 \end{array} & g_{\alpha\beta} \\ \end{array} \right)\;,
\end{equation}
i.e., the time-time component is just the flat one. Here $\mu, \nu= 0,1,2$ and $\alpha, \beta=1,2$, $q^\mu \equiv (t, u, v)$ where $t$ coincides with the laboratory time and $u,v$ are the coordinates on the surface. Second, we have to use a Lagrangian rather than a Hamiltonian description. Consequently, the dynamics of graphene's conductivity electrons near a Dirac point is given by the customary generalization to a curved spacetime\cite{wald, birrell} of the action for massless Dirac spinors in 2+1 dimensions
\begin{equation}\label{mainaction}
{\cal A} = i\int d^3 q \sqrt{g}\, {\bar \psi} (q) \gamma^\mu \nabla_\mu \psi (q)\,,
\end{equation}
where $\hbar = v_F = k_B = 1$, see \cite{note1} for conventions. We are building upon the continuum description of a Dirac quantum field, and upon modeling the effects of curvature through the coupling of the Dirac field to a curved \emph{spatial} metric. The effectiveness of both assumptions to describe graphene's quasiparticles dynamics was proven elsewhere\cite{vozmediano2,vozmediano}. Here we assert that the electrons on graphene might directly experience a curved space{\it time} even though the curvature is all in the spatial part. When the metric (\ref{mainmetric}) is conformally flat we can make use of the Weyl symmetry\cite{lor} enjoyed by the action (\ref{mainaction}) to obtain exact results that otherwise are difficult or impossible to obtain. As proved in earlier work\cite{iorio}, the metric (\ref{mainmetric}) is conformally flat for all surfaces of constant Gaussian curvature ${\cal K}$, but we need (\ref{mainmetric}) to \textit{explicitly} take the form $g^{\rm graphene}_{\mu \nu} ({\cal Q}) = \Phi^2 ({\cal Q}) g^{\rm flat}_{\mu \nu} ({\cal Q})$, for certain coordinates ${\cal Q}^\mu$. Only then we can use all the power of Weyl symmetry. Those coordinates surely exist, but the key issue for graphene is whether those ${\cal Q}^\mu$ can be \textit{practically realized in the laboratory}\cite{note3}.

\begin{figure}
\centering \leavevmode \epsfxsize=8.5cm \epsfysize=5cm
\epsffile{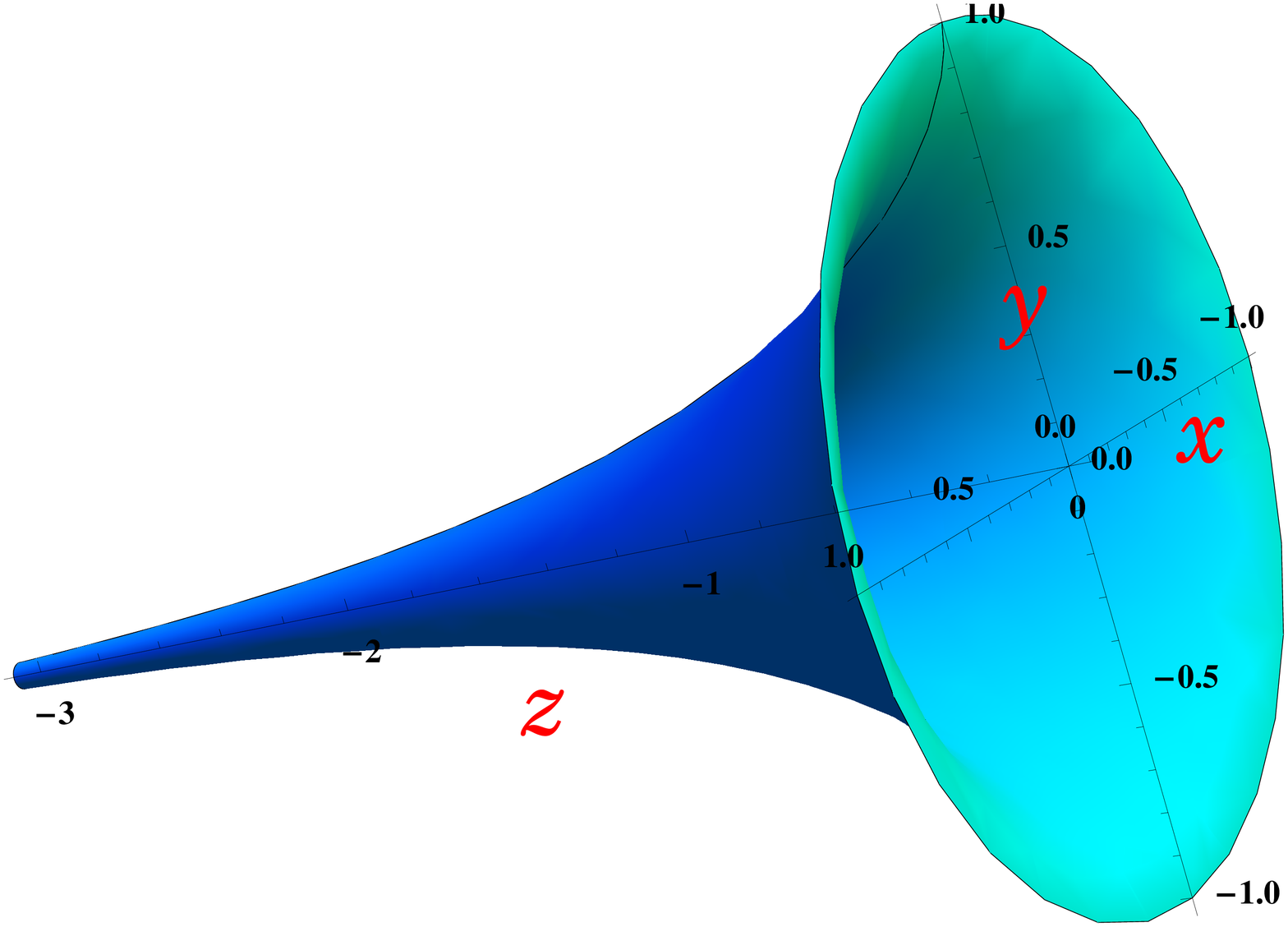}
\caption{}
\label{Beltrami}
\end{figure}

One of the main results of this work is that the Beltrami pseudosphere\cite{eisenhart}, $d{\ell}^2 = g_{\alpha \beta} q^\alpha q^\beta = du^2 + r^2 e^{2 u/r} dv^2$, with $v\in [0, 2\pi]$, $u\in [-\infty, 0]$ (see Fig.~\ref{Beltrami}), solves the problem. For this surface the coordinates ${\cal Q}^\mu$ are the coordinates $q^\mu$, and the time coordinate coincides with the laboratory time. To see it, we first write the spatial line element in isothermal coordinates $(\tilde{x}, \tilde{y})$ \cite{note3}, where we use the upper-half plane model of Lobachevsky geometry, valid for any surface of constant Gaussian curvature. Then it is immediate to write the whole line element, including time, in an explicitly conformally flat fashion \cite{note3}. The important point here is that, for the physical application of this non-Euclidean geometry, we have to be able to express the abstract coordinates $(\tilde{x}, \tilde{y})$ in terms of coordinates measurable within the real (Euclidean) space of the lab ${\bf R}^3$. For the Beltrami pseudosphere we have ${\tilde x}= v / r$ and ${\tilde y}= e^{-u/r} / r$, hence $ds^2_{\rm graphene} = e^{2u/r} \left[e^{-2u/r}(dt^2-du^2)-r^2dv^2\right]$. As the line element in square brackets is flat (and Rindler\cite{takagi,birrell}), for a Beltrami pseudosphere we can fulfill the condition of a physically doable $g^{\rm graphene}_{\mu \nu} (q) = \Phi^2 (q) g^{\rm flat}_{\mu \nu} (q)$ already in the frame $q^\mu$. From now on, the spacetime we shall suppose to be experienced by graphene's quasi-particles, is
\begin{equation}\label{beltramimetric}
    g^{(B)}_{\mu \nu} (q) = \left(
                    \begin{array}{ccc}
                      1 & 0 & 0 \\
                      0 & - 1 &  0 \\
                      0 & 0 & - r^2 e^{2 u/r} \\
                    \end{array}
                  \right) \,,
\end{equation}
where, $t \in [- \infty, + \infty]$, $u \in [- \infty, 0]$, $v \in [0,2\pi]$. We call this a ``Beltrami spacetime''. Once we take this view, any measurement is an operation relating different spacetimes: the inner (2+1)-dimensional curved spacetime and the outer (3+1)-dimensional flat spacetime. Let us explain how we approximate such a situation:

The electrons, far from the graphene surface, loose the lattice-induced properties of quasi-particles, such as the pseudo-relativistic Dirac nature hence the final ground state reached after the measurement is of a non-relativistic nature. To accommodate this hybrid situation into the fully relativistic scenarios of QFT in curved spacetimes, when we take into account the different quantum vacua for the different observers involved\cite{birrell,israel}, we approximate the ground state associated with the measurements with the Minkowskian one, $|0_M\rangle$. This way we have both, the mathematical description of the experimental evidences of the lattice-induced features (we keep a relativistic-like, (2+1)-dimensional structure), and the information on the true ground state (we use a flat vacuum defined everywhere). This is the best approximation: the Dirac field is living in a (2+1)-dimensional curved spacetime with coordinates $q^\mu \equiv (t, u, v)$, the measuring device has the same coordinates (i.e. it follows the profile of the surface in a specific manner described later), the quantum vacuum of reference entails information on the flatness of the ambient spacetime but retains the basic lattice-induced features.

We shall focus on the one particle Green's function that contains all the information on the single particle properties of the system such as the LDOS, life time of the quasi-particles and thermodynamic properties (specific heat). For the reasons illustrated above, this is defined as
$S^{(B)}(q_1, q_2)\equiv \langle 0_M|\psi^{(B)}(q_1){\bar \psi}^{(B)}(q_2)|0_M\rangle$, the positive frequency Wightman function, in the language of QFT in curved spacetimes\cite{birrell,takagi} (see also \cite{israel}). To obtain an exact result for this function, we shall use local Weyl symmetry as this case is a perfect match for its implementation\cite{iorio}: $g_{\mu\nu}^{(B)} = \varphi^2(u)  g_{\mu\nu}^{(R)}$, $\psi^{(B)} = \varphi^{-1}(u) \psi^{(R)}$, with $\varphi (u) = e^{u/r}$ and the metric $g^{(R)}_{\mu \nu} (q) = {\rm diag} (e^{-2u/r}, - e^{-2u/r}, - r^2)$,
describes a flat geometry. ``$R$'' here stands for ``Rindler''. Local Weyl symmetry gives ${\cal A}_B = i\int d^3 q \sqrt{g^{(B)}}\, {\bar \psi}^{(B)}(q) \gamma^\mu \nabla_\mu \psi^{(B)}(q)= i\int d^3 q \sqrt{g^{(R)}}\, {\bar \psi}^{(R)}(q) \gamma^\mu \nabla_\mu \psi^{(R)}(q)=A_R$, with ${\cal A}_B$ referring to a {\it curved} spacetime, and $A_R$ to {\it flat} curvilinear coordinates. The Green's function of interest can then be written as
$S^{(B)}(q_1, q_2)=\varphi^{-1}(q_1) \varphi^{-1}(q_2)S^{(R)}(q_1, q_2)$, where $S^{(R)} (q_1, q_2) \equiv \langle 0_M|\psi^{(R)}(q_1) {\bar \psi}^{(R)}(q_2)|0_M\rangle$. We shall now study the spacetime $g_{\mu\nu}^{(R)}$ to compute $S^{(R)}$, but one must bear in mind that this is a {\it fictitious} spacetime. The only physical spacetime and Green's function are the Weyl-equivalent $g_{\mu\nu}^{(B)}$ and $S^{(B)}$, respectively.

Let us introduce Minkowski coordinates $Q^\mu=(T, X, Y)$ for which  $ds_{(R)}^2=g_{\mu\nu}^{(R)}dq^\mu dq^\nu=\eta_{\mu\nu} dQ^\mu dQ^\nu$: $T=r e^{-u/r}\sinh \frac{t}{r}$, $X=rv$, $Y=r e^{-u/r}\cosh \frac{t}{r}$. Although no physical quantity explicitly depends from these coordinates, we have assumed that the quantum vacuum of reference is Minkowskian. We have that $Y^2-T^2=r^2 e^{-2u/r}\equiv \alpha^{-2}(u)$, which, for constant $u$, corresponds to worldlines of observers moving at constant ``proper acceleration''\cite{wald,birrell,takagi} $\alpha(u)\equiv e^{u/r}/r \in [0, r^{-1}]$. In our Rindler spacetime there is an unusual maximal ``acceleration'', $\alpha_{\rm max}\equiv\alpha(u=0)=r^{-1}=\sqrt{-{\cal K}}$, of complete geometric origin. Furthermore, we are forever confined to one Rindler wedge ($\alpha \geq 0$) and on one side ($u \leq 0$) of the singular maximal circle\cite{note3} (``Hilbert horizon''), the other side is unaccessible  as ``the world ends'' at $R=r$ which corresponds to $u=0$. For a radius of curvature $r \sim 1$mm, the Rindler horizon $Y=T$ is effectively reached after a time $t_{\rm hor} \sim r/v_F$ of few nanoseconds ($v_F \sim 10^6$m/s). Let us also introduce a perhaps more familiar notation $\eta = t/r = \alpha_{\rm max} t\,, \quad  \xi = r e^{-u/r}=\alpha^{-1}(u)$ with which $ds^2_{(R)}=\xi^2 d\eta^2-d\xi^2-r^2dv^2$. The real coordinates $q^\mu$ are those of an ``accelerated'' observer in the fictitious spacetime and constant ``acceleration'' means $u=constant\equiv{\bar u}$ (we also take $v=constant\equiv{\bar v}$). Thus, to fit our set-up within the truly relativistic requirements we need to refer to the Green's function $S^{(R)}$ at {\it the same point in space and at two different times} $S^{(R)}(t_1-t_2, {\bf q}, {\bf q}) \equiv \langle 0_M| \psi^{(R)}(t_1; {\bf q}) {\bar \psi}^{(R)}(t_2; {\bf q})|0_M\rangle$, as this is what would be seen by an observer of the above described worldline. The dependence on $t_1-t_2$ is a result of the stationarity of the worldline in point. To take into account the nonzero size of the detector we need to compute $S^{(R)}$ by setting $t\to t+i\varepsilon$ ($t_1-t_2\equiv t$)\cite{birrell, takagi}.
For the experiment we have in mind, $\varepsilon$ is the size, in ``natural units'', of the STM needle or tip. For a tungsten needle $\varepsilon \sim 0.25{\rm mm} \times v_F^{-1} \sim 10^{-10}$s, while for a typical tip $\varepsilon \sim 10 {\rm {\AA}} \times v_F^{-1} \sim 10^{-15}$s (see, e.g.,\cite{stmtip}). Hence, the Unruh requests are satisfied by considering the Green's function $S^{(R)}(\tau, {\bf q}, {\bf q})$, where $\tau \equiv t / e^{{\bar u}/r}$, and measuring at each point for a time given by the largest among $\varepsilon$ and the $t_{\rm hor}$ ($t_{\rm hor} \sim 10^{-9}$s for $r \sim 1$mm and $t_{\rm hor} \sim 10^{-12}$s for $r \sim 1\mu$m). The Unruh thermal features are then readily seen by considering, as customary, the power spectrum\cite{birrell,takagi}
\begin{equation}\label{fourierF}
    F^{(R)}(\omega, {\bf q})\equiv \frac{1}{2} \text{Tr}\left[\gamma^0\int_{-\infty}^{+\infty}d\tau e^{-i\omega \tau} S^{(R)}(\tau, {\bf q}, {\bf q}) \right]\,,
\end{equation}
that, for graphene, besides inessential constants, coincides with the definition of the electronic LDOS\cite{altland,vozmediano}, $\rho^{(R)}(\omega, {\bf q}) \equiv \frac{2}{\pi} F^{(R)}(\omega, {\bf q})$.

As we are in a massless model, $F^{(R)}$ can be computed exactly, i.e. without resorting to a perturbative expansion. To see it, first one recalls that on general grounds and for any spacetime dimensions $n$, the Dirac ($S_n$) and scalar ($G_n$) Green's functions are related as: $S_n =  \not\!\partial G_n$, see, e.g., \cite{birrell} for $m=0$. With our choice of the worldline (i.e., for us, of the measuring procedure) we then have the exact expression: $S_n^{(R)}(\tau) = \gamma^0 \partial_z G_n^{(R)}(\tau) = \lambda_n G_{n+1}^{(R)}(\tau)$, where $z = \varepsilon + 2 i \alpha^{-1} \sinh(\alpha \tau/2)$ and $\lambda_n = 2 \sqrt{\pi} \Gamma(n/2)/\Gamma((n-1)/2)$. Thus, by taking the Fourier transform and the trace, as in (\ref{fourierF}), one easily obtains $F_n^{(R)}(\omega) = \lambda_n D_{n+1}^{(R)}(\omega)$, where $D$ refers to the power spectrum for a {\it scalar field}. The expression for the latter is customary, and it is given by (see, e.g., par. 4.1 of \cite{takagi})
\begin{equation}\label{scalarnoiseN}
D_{n+1}^{(R)}(\omega) = \frac{2^{1-n} \pi^{(1-n)/2}}{\Gamma(n/2)} \frac{|\omega|^{n-1}/\omega}{e^{2 \pi \omega / \alpha}-(-1)^{n+1}} \;,
\end{equation}
where $n = 2,3,4$. By setting $n=3$, we then immediately obtain the expression we are looking for: $F^{(R)}(\omega, {\bf q}) = \frac{1}{2}\, \omega / (e^{\omega/{\cal T}}-1)$, where $\cal T$ is a Unruh temperature \cite{unruh}
\begin{equation}\label{36}
    {\cal T}\equiv \frac{\alpha({\bar u})}{2\pi}=\frac{e^{{\bar u}/r}}{2\pi r}\equiv {\cal T}_0 \, e^{{\bar u}/r}\,,
\end{equation}
that includes the constant ${\cal T}_0 = 1/ (2\pi r)$ and the Tolman factor\cite{wald} $e^{{\bar u}/r}$, as required by local measurements.

\begin{figure}
\centering \leavevmode \epsfxsize=9cm \epsfysize=5.5cm
\epsffile{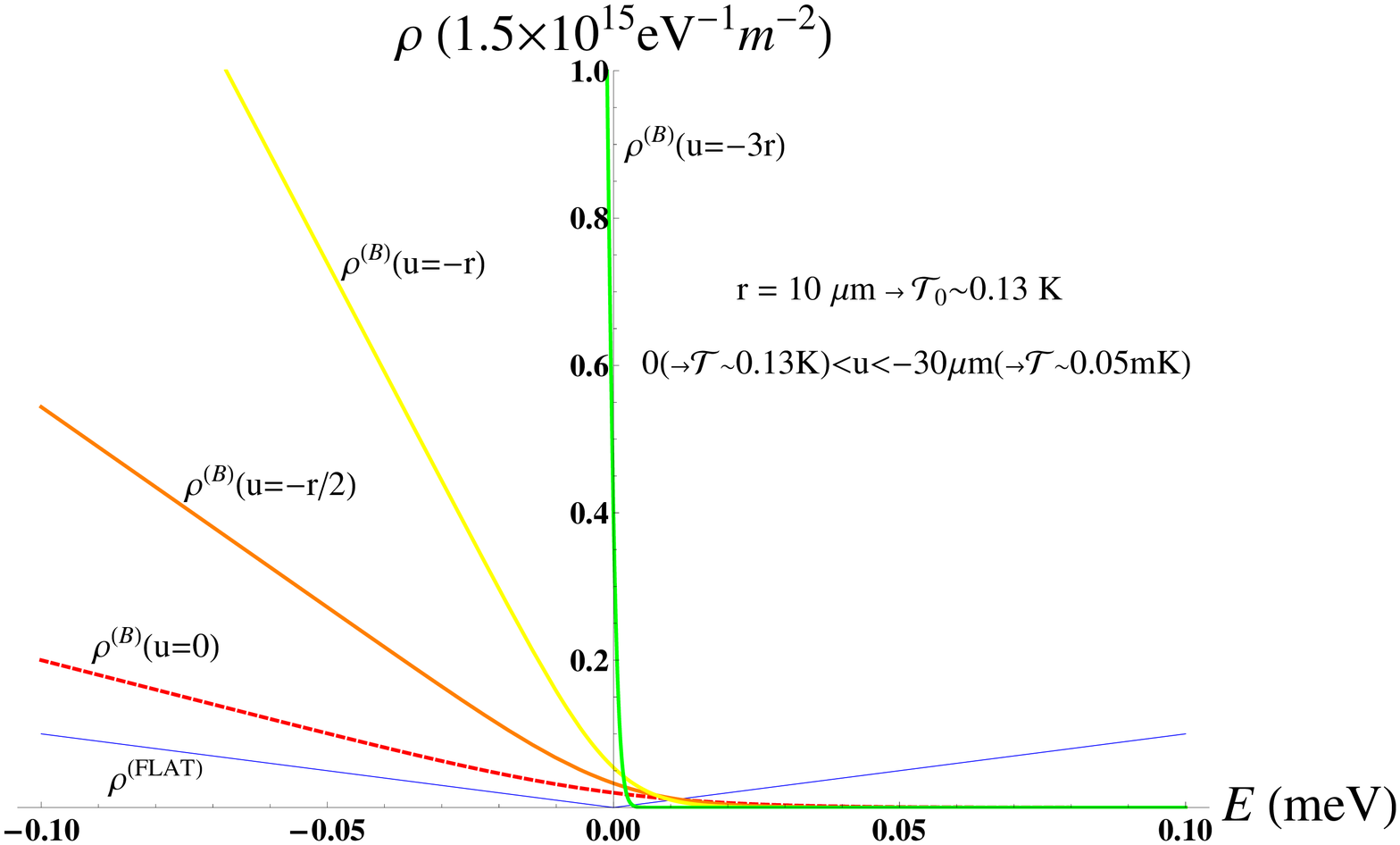}
\caption{}
\label{linecut}
\end{figure}

We see here a thermal distribution, of the Bose-Einstein kind (even though this is a Dirac system). The latter instance is known as ``statistics inversion''\cite{note2}. As for the thermal distribution {\it per se}, it must be seen as the consequence of a Unruh phenomenon taking place in the fictitious Rindler spacetime, to which the real Beltrami spacetime is related through the Weyl transformation. In that fictitious spacetime, the Unruh phenomenon is taking place exactly in the same way as for the standard derivation of Unruh \cite{unruh, birrell}: an accelerated observer (constant $u$), sees an event horizon, hence half of the modes (say the ``outgoing'') are inaccessible to measurements,  and tracing them away produces a thermal distribution.

This is {\it not} what is taking place on the real graphene sheet, as the physical result is only recovered {\it once we move to the Beltrami spacetime}. The marvelous thing with the symmetry we are exploiting (Weyl symmetry) is that the latter step is very simple, and preserves the key features of the Unruh phenomenon, although turning it into a Hawking phenomenon, due to the presence of curvature. Indeed, as the Weyl factor in $S^{(B)}$ is time-independent, it goes through the Fourier transform, i.e. $F^{(B)} = \varphi^{-2}({\bf q}) F^{(R)}$, with obvious notation, hence the predicted physical LDOS is $\rho^{(\text{B})}(\omega, {\bf q}) = \varphi^{-2}({\bf q}) \rho^{(R)}(\omega, {\bf q})$. More explicitly
\begin{equation} \label{37}
    \rho^{(\text{B})}(E, {\bar u}, r)= \frac{4}{\pi} \frac{1}{(\hbar v_F)^2} \frac{E \; e^{-2{\bar u}/r}}{\exp{\left[ E / (k_B {\cal T}_0 e^{{\bar u}/r}) \right]}-1} \;,
\end{equation}
where we included the $g=4$ degeneracy, and the proper dimensional units are re-introduced, for instance $\omega \equiv \omega / v_F$, $E \equiv \hbar \omega$ and ${\cal T}_0 \equiv \hbar v_F / (k_B 2 \pi r)$. With ${\bar u}$ we stress that: i) the measuring device has to closely follow the profile of the surface, so that its coordinates can be taken to be $q^\mu$; and ii) the device has to stop at each given point on the surface $({\bar u}, {\bar v})$ for a time much bigger then max$(\varepsilon, t_{\rm hor})$, not a stringent condition, as we saw earlier. This Hawking effect is inferred from an equivalent (through Weyl symmetry) Unruh phenomenon. Also in \cite{deser} the Hawking effect is obtained from a Unruh effect, but due to our choice of the embedding those results differ from ours.

To measure the effect with an STM device we need to follow the prescriptions i) and ii) above and, by fixing the right polarity of the bias voltage, we need the tunneling current to be that of the sample electrons tunneling to the tip, and not viceversa. The plot we need is a current-to-voltage ($dI/dV$) spectrum obtained by first constructing a one-layer graphene sample shaped as a Beltrami pseudosphere with fixed $r$ and by measuring at a particular fixed value of $u$. We then vary the bias voltage to obtain the $\rho^{(\text{B})}$ vs $E$ behavior. As the formula (\ref{37}) does not depend on the angle $v$, we do not need to stay at a fixed point $(u,v)$ on the surface but we can loop around the circle at fixed $u$ (see Fig.~1) while varying the bias voltage $V$. We can then repeat the same measurements at different values of $u$ to obtain a series of $dI/dV$ spectra (a ``line-cut``). The expected line-cut is illustrated in Fig.~2 for $r = 10\mu$m.

Our model has an intrinsic energy scale, $E^* = \hbar v_F/r \simeq 6.6 \times 10^{-7} {\rm eV} \times [r \; {\rm mm}]$, induced by the radius of curvature. Only electrons with an energy below this threshold will have a long enough wave-length to experience the whole pseudosphere, hence their contribution to the LDOS will be appreciated by our model. Another way of saying the same thing, is that only electrons with a small enough energy, on the scale of $E^*$, will have enough time to travel back and forth from the given point $u$ on the pseudosphere to contribute to our $\rho^{(\text{B})}(u)$. This means that our analysis can only predict the behavior till $E^*$, while beyond that the curve for the LDOS may indeed rise-up again to include electrons with positive energies that do not appear within the range considered in Fig.~2. Thus, such electrons have not disappeared from the spectrum, but rather there has been a re-arrangement that pushed them beyond $E^*$.

To avoid defects proliferation\cite{kleinert} we take small curvatures on the natural scale of the lattice spacing $\ell \sim$\AA $\;$: $r >> \ell$. We take $r \sim 10^{-7}$m as the highest curvature (smallest value of $r$), hence ${\cal T}_0 \sim 13$K is the highest detectable temperature. The zero-curvature limit (large $r$) of $\rho^{(\text{B})}$ does not match the flat LDOS, $\rho^{(\text{flat})}(E) = \frac{2}{\pi} \frac{1}{(\hbar v_F)^2} |E|$. This is as it must be since (\ref{37}) is not the result of a perturbative computation with $\rho^{(\text{flat})}(E)$ as the leading term, thus even a very small curvature fully turns on the effect. The limit for zero energy gives a nonzero result, $\rho^{(\text{B})}(0, {\bar u}, r) = \frac{2}{\pi^2} \frac{1}{\hbar v_F} \, \frac{e^{-{\bar u}/r}}{r}$, as expected for a finite temperature LDOS. A direct measurement of this value of the LDOS would give the constant Hawking temperature ${\cal T}_0 = \frac{\pi}{4} \frac{(\hbar v_F)^2}{k_B} e^{{\bar u}/r}\rho^{(\text{B})}(0, {\bar u}, r)$.  Furthermore, the Hawking temperature can be measured for each power spectrum (fixed $u$), giving the corresponding $u$-dependent temperature, ${\cal T} = E/ [k_B \ln(\frac{4 E e^{-2 \bar{u}/r}}{\pi (\hbar v_F)^2 \rho^{(\text{B})}} + 1)]$. Different ${\cal T}$s for different $u$s will then give the same constant Hawking temperature through ${\cal T}_0 = e^{-u/r} {\cal T}$.

{\it Acknowledgements.} We thank A.~MacDonald for clarifying the importance of the energy scale $E^*$, and N.~Hitchin for help with the geometry of the elliptic pseudosphere. We also thank S.~Bonanos, I.~Brihuega, F.~Guinea, and M.~Vozmediano.

\newpage    

\noindent {\bf Figure captions}

\bigskip

\noindent Caption for Figure 1:

{\bf Beltrami pseudosphere.} The ${\bf R}^3$ coordinates of the Beltrami pseudosphere, in the canonical form, are\cite{eisenhart} $x(u,v) = R(u) \cos v$, $y(u,v) = R(u) \sin v$, $z(u) = r (\sqrt{1 - R^2(u)/r^2} - {\rm arctanh}\sqrt{1 - R^2(u)/r^2})$, with $R(u) =  c \, e^{u/r}$, $c > 0$ and $r = \sqrt{-{\cal K}^{-1}} > 0$ where ${\cal K}$ is the constant negative Gaussian curvature.  In this paper we choose $c=r$, thus $R(u) \in [0,r]$ as $u \in [-\infty, 0]$. The surface is not defined for $R>r$ ($z$ becomes imaginary). In the plot $r=1$ and $u \in [-3.37, 0]$, $v\in [0, 2\pi]$.

\bigskip

\noindent Caption for Figure 2:

{\bf The expected LDOS.} Series of $dI/dV$ spectra (``line-cut'') from an STM. The model applies to energies below the natural scale, $E^* = \hbar v_F/r \simeq 6.6 \times 10^{-7} {\rm eV} \times [r \; {\rm mm}]$, where $[r \; {\rm mm}]$ is the numerical value of $r$ measured in mm. In this plot $E \in [-0.1,0.1]$meV, corresponding to $r = 10\mu$m. The flat LDOS is shown in blue. $u \in ]0, -3r]$, $u=0$ (red dashed curve) corresponds to the singular boundary of the pseudosphere, beyond the upper limit $u=-3r$ (green curve) the pseudosphere is too sharp. To each spectrum corresponds a $u$-dependent temperature $\cal T$ while the constant temperature, same for the whole line-cut, is ${\cal T}_0 = e^{-u/r}{\cal T} \sim 0.13$K.


\begin{thebibliography}{}
\bibitem{note1} The indices $\mu , \nu = 0, 1,2$ respond to diffeomorphisms (Einstein indices), while $a, b = 0, 1,2$ respond to flat space transformations (Lorentz indices), $\eta_{a b} = {\rm diag}(+1, -1, -1)$, $\sqrt{g}=\sqrt{\det g_{\mu \nu}}$ and the diffeomorphic covariant derivative is $\nabla_\mu=(\partial_\mu + \frac{1}{2} \omega_\mu^{\; b c} J_{b c})$, with $J^{a b} =  \frac{1}{4} [\gamma^a, \gamma^b]$, and ${\omega_\mu}^a_{\; b} = e^a_\lambda (\delta^\lambda_\nu \partial_\mu + \Gamma_{\mu \nu}^\lambda) E^\nu_b$
is the spin connection, where $\Gamma_{\mu \nu}^\lambda$ is the Christoffel connection, $e^a_\mu$ the Vielbein and $E_a^\mu$ its inverse, $\eta_{a b} e^a_\mu e^b_\nu = g_{\mu \nu}$, $e^a_\mu E_a^\nu = \delta_\mu^\nu$, $e^a_\mu E_b^\mu = \delta_b^a$. Our Riemann tensor is ${R^\rho}_{\lambda \mu \nu} = \partial_{[\nu} {\Gamma^\rho}_{\mu] \lambda} + {\Gamma^\rho}_{[\nu \sigma} {\Gamma^\sigma}_{\mu] \lambda}$, torsion $T^\lambda_{\;\; \mu \nu} = \Gamma^\lambda_{\;\; [ \mu  \nu]}$ is zero. For more expressions see \cite{iorio}.
\bibitem{note2} This is quite a general feature of the power spectrum (not of the expectation value of the number operator!), whose physical origins are not understood as yet, but that clearly depends upon the alternating sign in (\ref{scalarnoiseN}), i.e. on the dimensions of the spacetime. See \cite{takagi} and also \cite{hyun1}.
\end{thebibliography}
\end{document}